\documentclass[final]{aipproc}
\layoutstyle{8x11double}
\usepackage{graphicx}
\usepackage{graphics}

\begin{document}

\title{Fluctuations in Student Understanding of Newton's 3rd Law}

%01.40.Fk 	Research in physics education, 01.30.Cc 	Conference proceedings, 01.30.lb 	Undergraduate schools

\author{Jessica W. Clark}{
  address={Department of Physics and Astronomy, University of Maine, Orono, ME 04469}
}

\author{Eleanor C. Sayre}{
  address={Wabash College, Crawfordsville, IN 47933}
}

\author{Scott V. Franklin}{
  address={Department of Physics, Rochester Institute of Technology, Rochester, NY 14623}
}

\begin{abstract}
We present data from a between-student study on student response to
questions on Newton's Third Law given throughout the academic year.
The study, conducted at Rochester Institute of Technology, involved
students from the first and third of a three-quarter sequence.
Construction of a response curve reveals subtle dynamics in student
learning not captured by simple pre/post testing.  We find a a
significant positive effect from direct instruction, peaking at the
end of instruction on forces, that diminishes by the end of the
quarter.  Two quarters later, in physics III, a significant dip in
correct response occurs when instruction changes from the vector
quantities of electric forces and fields to the scalar quantity of
electric potential.  Student response rebounds to its initial values,
however, once instruction returns to the vector-based topics involving
magnetic fields.
\end{abstract}

\maketitle

\section{Introduction}

Pre/post testing of students is virtually the standard for assessing
learning in physics \cite{TwoCurr}.  Thornton and Sokoloff \cite{FMCE}
used the method to establish the validity of the FMCE and demonstrate
the efficacy of active engagement classrooms, a study reproduced on a
much larger scale by Hake \cite{IEvsT}.  Pre/post testing fails,
however, to reveal the rich dynamism of student learning.  Research on
forgetting and interference \cite{CriticalIssues, LongTerm,
  ClassicalCond,ContextTime} shows that learning is very subtle, and
often time-dependent, with even significant gains sometimes short
lived.  Sayre and Heckler have applied the between-student method
\cite{PeaksDecay} to physics classes.  In this method, data are
collected regularly through the academic term with comparison between
different groups of students.  While this requires significantly
larger student populations --- and overhead on grouping --- when
successful it allows for a much more detailed picture of student
understanding before, during, and after instruction.

\section{Methods}
\subsection{Population}

RIT is classified \cite{Carnegie} as a large four-year, private
university with high undergraduate enrollment, balanced arts \&
sciences/professions, and some graduate coexistence.  At the time of
this study, the academic year was divided into four 10-week quarters
(including the summer term).  Each year $\approx 2000$ students take
introductory calculus based physics, which is offered in a workshop
format that integrates lecture, experiment, and short group
activities.  Adapted after the SCALE-UP project \cite{scaleup}, the
classes meet for three 2-hour sessions each week, with students seated
at tables of six and working in small groups.  Classrooms accommodate
up to forty-two students, with enrollment in each section varying.

\begin{table}
	\centering
		\begin{tabular}{|l|c|c||c|c||c|}\hline
		\multicolumn{1}{|c}{ } & \multicolumn{2}{|c||}{Fall} &\multicolumn{2}{c||}{Winter} & \multicolumn{1}{c|}{\bf{Total}} \\\hline
			Course & \# of Sec. & N & \# of Sec. & N & \bf{N} \\\hline
			Physics I & 5 & 142 & 14 & 441 & \bf{583} \\\hline
			Physics III & 8 & 257 & 5 & 144 & \bf{401} \\\hline
		\end{tabular}	
	\caption{Study population for fall and winter divided by course.  In fall, the primary course is physics III and in winter, the primary course is physics I.}
	\label{tab:pop}
\end{table}

Students are initially sorted into two different tracks based on an
Institute math placement exam.  Students who fail to achieve a minimum
score on this exam are placed into slower-paced calculus and physics
sequences.  These students meet for four 2-hour workshop meetings
(instead of three).  In order to ensure that the same material is
covered in the regular and remedial sequences, a mid-term and final
exam are common across all sections.  These comprise 60\% of the
students' final grades.  Students who achieve an A in the remedial
sequence may enroll in the mainstream sequence for subsequent
quarters.  All students are ``mainstreamed'' by the third quarter for
Electricity and Magnetism.  In E\&M, there are three tests and a
final.  All tests are common across sections, with the final all
multiple choice.  Multiple sections of Mechanics and E\&M are offered
every quarter, with most students beginning the sequence in the Winter
of their freshman year.

Our study tested students in the first (Mechanics) and third (E\&M)
quarters during the fall and winter quarters of the 2009-2010 academic
year.  In the fall there were 5 sections of Mechanics (three
mainstream ($N=109$) and two remedial ($N=33$)) and eight sections of
E\&M ($N=257$).  In the winter, there were fourteen sections of
Mechanics (eight mainstream ($N=281$) and six remedial ($N=160$)), and
five sections of E\&M ($N=144$).  We found little significant
difference between mainstream and remedial sections on the questions
involved in this study; henceforth we group the two together for
improved statistics.  Engineering students dominate the population,
comprising 57-83\% of the students in Mechanics and 65-83\% of E\&M
students.  Participation for each quarter is summarized in Table
\ref{tab:pop}.

\subsection{Between Student Testing}

RIT's system of multiple sections makes it ideal for a between-student
study.  First used in physics by Sayre and Heckler \cite{PeaksDecay},
this method gives short conceptual quizzes to different sub-groups of
the population in successive weeks.  This avoids test-retest effects
that would occur if the same group took the quiz twice.  Different
sections are therefore different groups, with section I (for example)
completing the relevant quiz in week 3 and section II taking the quiz
in week 8.  For this type of analysis to be valid, the groups must be
approximately normally distributed and have the same variance, which
we have confirmed.  The order in which each section took a quiz was
randomly assigned, and all quizzes were administered at the beginning
of class, once per week, in paper format.  Students had five-ten
minutes to complete each task, which were sometimes appended to an
instructor-generated quiz.

Because sections are statistically independent, we can compare the
performance of different sections across weeks, essentially capturing
student understanding on a weekly time scale.  A time plot of average
performance, termed the {\it response curve}, is sensitive to the
particulars of the week--- the current topic of instruction and
coincidence with exams or homework.  The conventional pre/post test
corresponds to the first and last points on the curve, and can miss
much of the dynamic evolution of understanding.  Error bars on the
response curve are determined using a binomial distribution, and we
collapsed data across the fall and winter quarters (by week) to
increase sample size.  Because the syllabus is unchanged, this is
appropriate.  Students in, for example, Week 3 in the Winter see the
same activities, labs, and lectures as those in Week 3 in the Fall.

\subsection{Tasks}

\begin{figure}
\centering
1	\includegraphics[width=.45\textwidth]{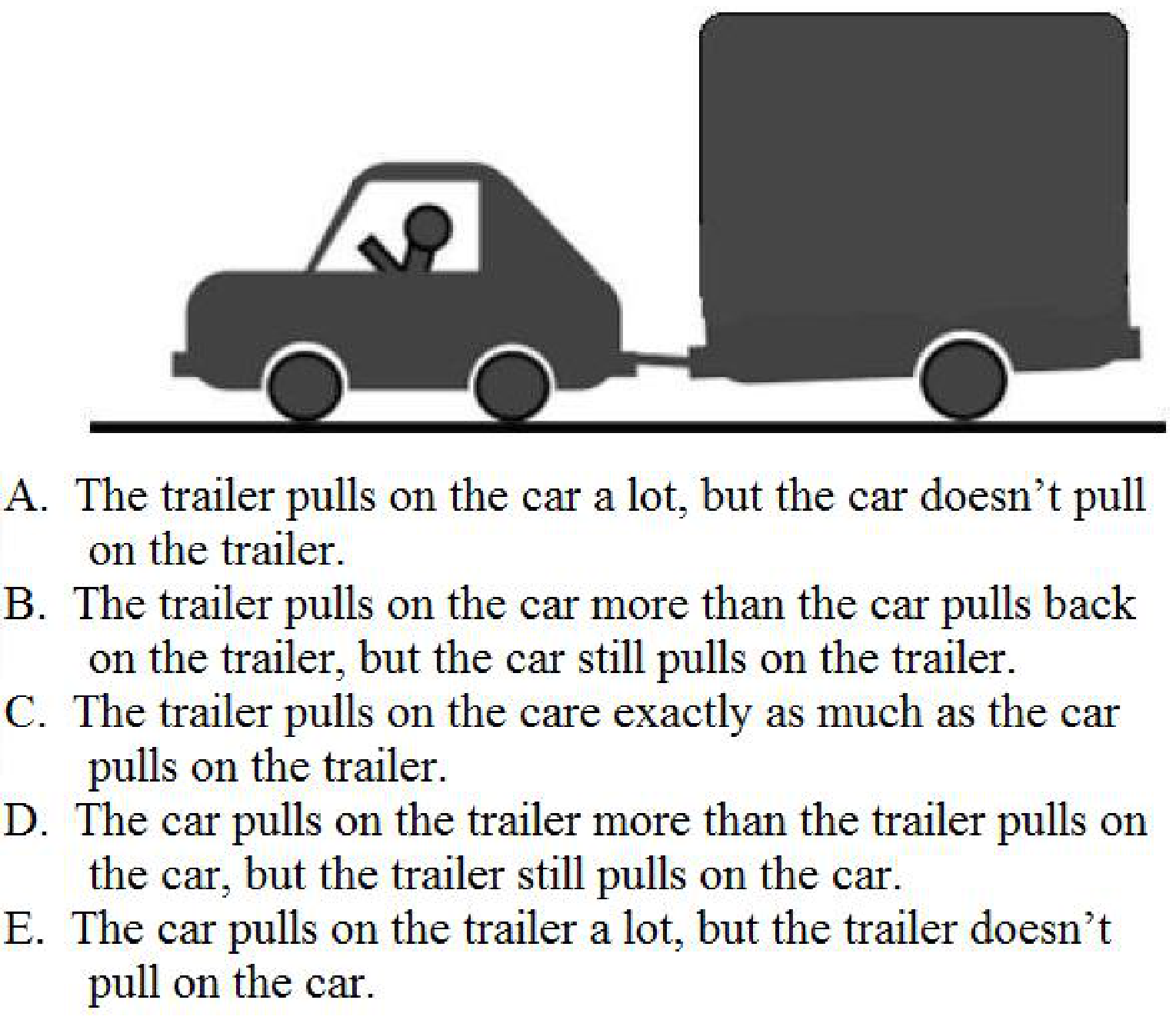}
	\caption{Prompt and responses for Newton's Third Law task for mechanics.  The students were asked to consider situations where the car was speeding up, constant speed up a hill, constant speed a level road, and slowing down.}
	\label{fig:mech}
\end{figure}

Tasks were devised to probe student understanding of Newton's Third
Law.  In order to align with the instruction of the different classes,
tasks were couched in appropriately different contexts.  (This also
made it easier for us to gain access to E\&M sections, as instructors
could more easily justify inclusion of the task.)  In Mechanics, the
task involved a car pulling a trailer (see Fig.~\ref{fig:mech}).
Students are asked to compare the forces acting on the car and trailer
as the car speeds up, travels at constant speed up a hill, travels at
constant speed on a level road, and slows down.  The answer choices
for each question were the same and the students could select each
answer as many times as they wanted.

\begin{figure}
	\centering
		\includegraphics[width=0.45\textwidth]{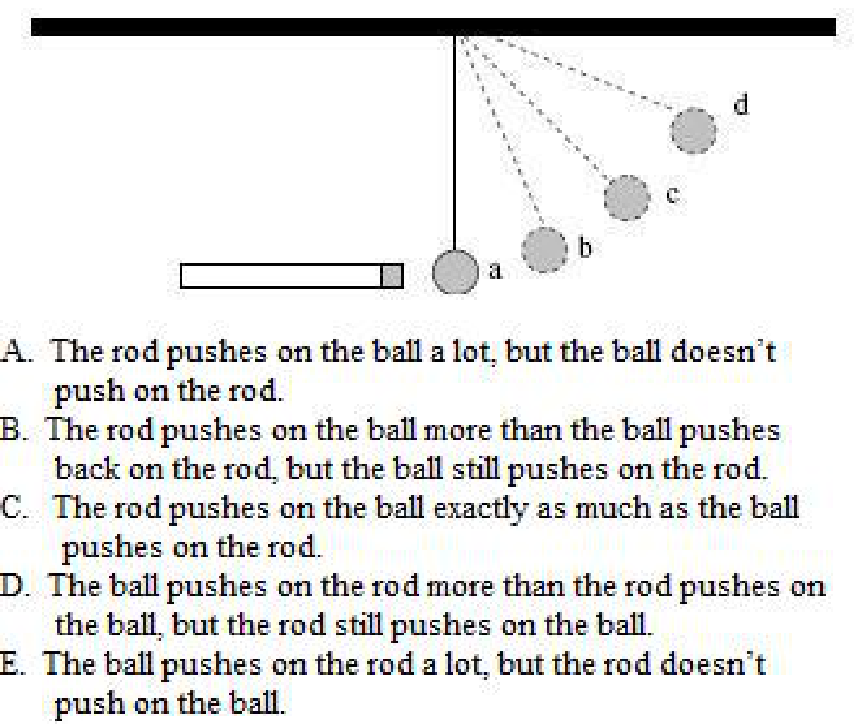}
	\caption{Prompt and response for Newton's Third Law task for physics III.  The students were asked to compare the forces on the rod and ball ball started to move, sped up, slowed down and at rest at the apex of the swing.}
	\label{fig:elec}
\end{figure}

For E\&M, the task was re-written to involve electric charges (see
Fig.~\ref{fig:elec}).  Students compared the forces acting on the rod
and ball as the ball starts to move, speeds up, and slows down as it
swings away from the rod and finally when it comes to rest.  This
question is not completely isomorphic to the Mechanics
formulation, and so we do not directly compare the Mechanics and E\&M
responses.  Rather, we look for changes in the response over the
course of each quarter, and similarities in how this behavior
corresponds to the topic of instruction.

\section{Results}
\subsection{Instruction's positive impact}

\begin{figure}
         \centering
                 \includegraphics[width=0.45\textwidth]{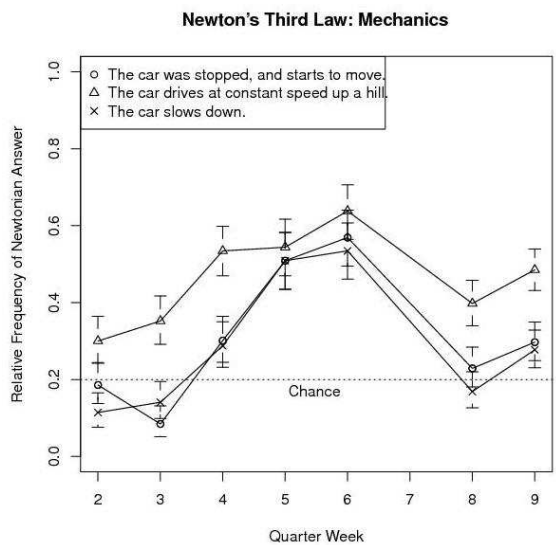}

         \caption{Response curve for physics I.  Before instruction response could be chance.  There is a broad peak during instruction with a maximum during week six which is the end of the section on forces.}

         \label{fig:n3lmechresp}
 \end{figure}

Figure \ref{fig:n3lmechresp} shows the response curve for students in
the Mechanics course.  (As noted above, responses are averaged across
all sections in both Fall and Winter.  A common assessment on an
unrelated topic was given to all sections in Week 7, so that data
point does not exist.)  Shown are average response for the three
non-trivial questions; the question involving the car traveling at
constant speed on level ground shows a ceiling effect where almost all
students answer correctly independent of week, instructor, or any
other variable.  Although 80\% of students have had physics prior to
the introductory course at RIT, response during the first few weeks of
the course, before explicit instruction of forces or Newton's Laws,
hovers around the chance line of 20\%.  Instruction on forces begins
in week 4, and student performance begins to rise, culminating with a
maximum performance in week 6.  Not coincidentally, week 6 is also the
last week of instruction on forces, and includes the examination.
After instruction, the response rapidly drops, with two of the
questions ending just above the chance line at the end of the quarter.
The small dip in week 8 may be due to instructor effect, but without
data from Week 7, we are unable to definitively identify a source for
the lower score.  The low score does not, however, negate the overall
result of rise and fall throughout the quarter.

 \subsection{Instruction's negative impact}

\begin{figure}
         \centering
                 \includegraphics[width=0.45\textwidth]{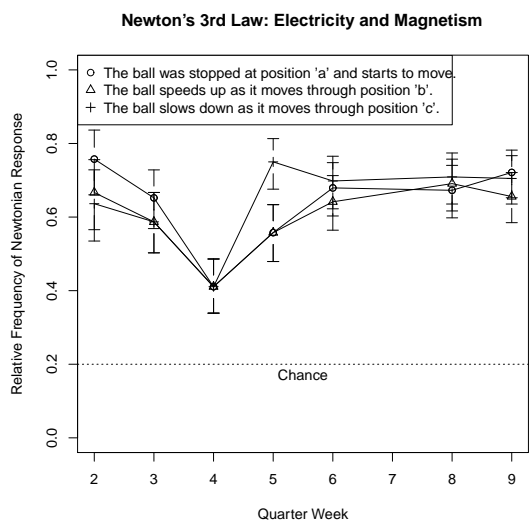}
         \caption{Response curve for physics III.  The response is mostly flat around the average of 66 percent with a measurable dip during week 4.  This dip corresponds to the period of instruction in electric potential.}
         \label{fig:n3lelecresp}
 \end{figure}

 Figure \ref{fig:n3lelecresp} shows the response curve for all students
 in the E\&M course, the question involving the ball at rest having
 been omitted due to the presence of a ceiling effect.  At RIT, E\&M is
 typically taken in the Fall quarter.  This means that because of the
 summer break, it has been approximately {\it 5 months} since these
 students last saw instruction on forces and Newton's laws.  (The
 Spring quarter deals with rotational motion, waves, and miscellaneous
 physics topics).  Nevertheless, students enter with an initial
 response of 66\%, significantly higher than they exited Mechanics.
 This has two potential explanations.  Most likely is a winnowing
 effect, with the weakest students leaving the sequence before reaching
 E\&M.  Failure rates (defined as obtaining a D, F, or withdrawing) in
 Mechanics average around 25\%, and an additional $\approx 17\%$ exit
 between Mechanics and E\&M.  Therefore, students entering E\&M are the
 top 62\% of the students in Mechanics, and a higher performance is
 expected.  Less likely is the possibility that instruction in the
 intermediate quarter has bolstered student understanding.  Subsequent
 testing in this course will take place next year to test this
 possibility.

 The most significant feature of Fig.~\ref{fig:n3lelecresp} is the
 pronounced dip in week 4 to 41\%.  This drop, 25\% points below the
 average, cannot be explained by instructor or section variance, and so
 we assert that course topic is the most likely cause.  In E\&M, the
 first three weeks are spent on electric fields, Coulomb's Law, and
 Gauss' law.  Week 4 shifts the topic from vector-based concepts to the
 scalar topics of electric potential and voltage.  We speculate,
 therefore, that instruction on the scalar electric concepts interferes
 with response to a vector-based (Coulombic force) question.  In week 5
 instruction shifts to current, resistance, and circuits.  While this
 is also scalar-based, and we note that the week 5 performance is still
 below average, we suspect that because instruction is not explicitly
 involving electric charges the interference effect is lessened.

 \section{Conclusions}

It has been established \cite{PeaksDecay} that student understanding
is dynamic, and time-dependent.  In this study we have shown that this
dynamism continues far beyond the immediate period surrounding
instruction.  Student response to questions on vector-based topics,
like Newton's Third Law, are sensitive to {\it any} physics
instruction they are receiving at the time.  ``Dissonant''
instruction, e.g. topics that emphasize a scalar concept, suppresses
student scores.  It is fortunate that this interference disappears
when instruction returns to more ``consistent'', i.e. vector-based,
topics.  We would expect a similar interference effect to occur in the
Mechanics quarter during instruction on the scalar topics of Energy.
Unfortunately that occurs during Weeks 7 and 8; we have no data for
Week 7 (due to a common assessment) and so we are unable to determine
if the low score in Week 8 is an instructor effect or interference.
Future work is looking into this question.

The impact of current instruction on previously learned knowledge has
been loosely termed ``interference''. \cite{ContextTime} It
underscores the complexity of student learning, as students struggle
to identify, activate, and use appropriate knowledge in response to a
prompt.  Even strong students, who have already progressed through two
previous quarters of physics and show a high initial score, struggle
to reconcile a strange prompt with their current frame of mind.  The
implications for testing and assessment may be profound, calling into
question the accuracy of any single evaluation.  It broadens the range
of contexts that are known to affect student performance (both
positively and negatively), and therefore cautions instructors from
reading too much into any single assessment.  Subsequent research will
look at interference effects in strong and weak students, mainstream
and remedial sections, and in more explicit vector tasks.

\begin{theacknowledgments}
We thank Gordon Aubrecht and Andrew Heckler for their assistance in
task development.  This work was originally carried out to satisfy
requirements for the RIT Physics Department Capstone Research program,
and we acknowledge the department for travel support.  This work is
partially supported by NSF DUE \#0941889 and \#0941378.
\end{theacknowledgments}

\bibliographystyle{aipproc}   % if natbib is available

\bibliography{Citation}

\begin{thebibliography}{10}
\expandafter\ifx\csname natexlab\endcsname\relax\def\natexlab#1{#1}\fi
\providecommand{\enquote}[1]{``#1''}
\expandafter\ifx\csname url\endcsname\relax
  \def\url#1{\texttt{#1}}\fi
\expandafter\ifx\csname urlprefix\endcsname\relax\def\urlprefix{URL }\fi

\bibitem[Kohlmyre et~al.(2009)]{TwoCurr}
Kohlmyre, M.~A., Cabellero, M.~D., Catrambone, R., Chabay, R.~W., Ding, L.,
  Haugan, M.~P., Marr, M.~J., Sherwood, B.~A., and Schatz, M.~F., \emph{Phys.
  Rev. ST Physics Ed. Research}, \textbf{5} (2009).

\bibitem[Thornton and Sokoloff(1998)]{FMCE}
Thornton, R.~K., and Sokoloff, D.~R., \emph{Americal Journal of Physics},
  \textbf{66}, 338--353 (1998).

\bibitem[Hake(1998)]{IEvsT}
Hake, R.~R., \emph{American Journal of Physics}, \textbf{66} (1998).

\bibitem[Postman and Underwood(1973)]{CriticalIssues}
Postman, L., and Underwood, B.~J., \emph{Memory and Cognition}, \textbf{1},
  19--40 (1973).

\bibitem[Semb et~al.(1993)]{LongTerm}
Semb, G.~B., Ellis, J.~A., and Araujo, J., \emph{Journal of Educational
  Psychology}, \textbf{85}, 305--316 (1993).

\bibitem[Rescorla and Wagner(1971)]{ClassicalCond}
Rescorla, R.~A., and Wagner, A.~R., \enquote{A theory of Pavlovian
  Conditioning}, in \emph{Classical Conditioning II: Current Theory and
  Research}, edited by A.~H. Black and W.~F. Prokasy, Meredith Corporation,
  1971.

\bibitem[Bouton(1993)]{ContextTime}
Bouton, M.~E., \emph{Psychological Bulletin}, \textbf{114}, 80--99 (1993).

\bibitem[Sayre and Heckler(2009)]{PeaksDecay}
Sayre, E.~C., and Heckler, A.~F., \emph{Physical Review Special Topics -
  Physics Education Research}, \textbf{5} (2009).

\bibitem[Car(2010)]{Carnegie}
 (2010), \urlprefix\url{http://classifications.carnegiefoundation.org/}.

\bibitem[Beichner et~al.(2000)]{scaleup}
Beichner, R.~J., Saul, J.~M., Allain, R.~J., Deardorff, D.~L., and Abbott,
  D.~S., Introduction to scale-up: Student-centered activities for large
  enrollment university physics (2000), presented at the Annual Meeting of the
  American Society for Engineering Education, Seattle, Washington.

\end{thebibliography}

\end{document}